\documentclass[11pt,twoside]{article}
\usepackage[english]{babel}
\usepackage{times,subeqnarray}
\usepackage{url}
\newif\myifpdf
\ifx\pdfoutput\undefined
   \usepackage[dvips]{graphicx}
\else
   \pdfoutput=1        % we are running PDFLaTeX
   \usepackage[pdftex]{graphicx}
\fi
\usepackage{apatitlepages}
\usepackage{psydraft}
\usepackage[natbibapa]{apacite}   % natbib
\myifpdf
  \DeclareGraphicsExtensions{.pdf,.eps,.png,.jpg,.mps,.tif}
\fi

\def\myheading{ Structure of Systematicity }

\begin{document}
\bibliographystyle{apacite}

% sloppy is the way to go!
\sloppy
\raggedbottom

\def\mytitle{ The Structure of Systematicity in the Brain }

\def\myauthor{Randall C. O'Reilly, Charan Ranganath \& Jacob L. Russin\\
  Departments of Psychology and Computer Science\\
  Center for Neuroscience\\
  University of California, Davis \\
  1544 Newton Ct\\
  Davis, CA 95618\\
  {\small oreilly@ucdavis.edu}\\}

\def\mynote{Submitted to Current Directions in Psychological Science.\\

  We thank Jon Cohen, Alex Petrov, Erie Boorman, Ignacio Saez, Taylor Webb, Stephen Frankland, James Antony, John Rohrlich, Mary Zolfaghar, and members of the CCN, DML, and Cohen labs for discussion and comments.

  R. C. O'Reilly is Chief Scientist at eCortex, Inc., which may derive indirect benefit from the work presented here.

  Supported by: ONR grants N00014-20-1-2578, N00014-19-1-2684/ N00014-18-1-2116, N00014-18-C-2067, N00014-17-1-2961, N00014-15-1-0033.  
}

\def\myabstract{
A hallmark of human intelligence is the ability to adapt to new situations, by applying learned rules to new content (systematicity) and thereby enabling an open-ended number of inferences and actions (generativity). Here, we propose that the human brain accomplishes these feats through pathways in the parietal cortex that encode the abstract structure of space, events, and tasks, and pathways in the temporal cortex that encode information about specific people, places, and things (content). Recent neural network models show how the separation of structure and content might emerge through a combination of architectural biases and learning, and these networks show dramatic improvements in the ability to capture systematic, generative behavior. We close by considering how the hippocampal formation may form integrative memories that enable rapid learning of new structure and content representations.
}

% \titlesepage{\mytitle}{\myauthor}{\mynote}{\myabstract}
% \twocolumn

%\titlesamepage{\mytitle}{\myauthor}{\mynote}{\myabstract}

\titlesamepageoc{\mytitle}{\myauthor}{\mynote}{\myabstract}

% single-spaced table of contents, delete if unwanted
% \newpage
% \begingroup
% \parskip 0em
% \tableofcontents
% \endgroup
% \newpage

% \twocolumn

\pagestyle{myheadings}

The ability to directly transfer existing knowledge to new situations by applying learned rules to new content (\emph{systematicity}) and thereby generate an open-ended number of different behaviors (\emph{generativity}) is particularly advanced in humans relative to other animals. However, this ability is nevertheless shared in a limited but instructive way with even relatively simple computer programs.  For example, it is easy to write a function that takes two arbitrary strings as arguments, and systematically returns the concatenation of those two strings. With just a bit more logic, calling such a function repeatedly can easily construct an open-ended number of sentences.  However, learning to do something like this in a neurally-plausible manner is much more challenging, and neural network models continue to be criticized as lacking in these signature human abilities, with much of their recent successes potentially attributable to something closer to rote memorization of increasingly large datasets \citep{FodorPylyshyn88,LakeBaroni17,Marcus18,OReillyPetrovCohenEtAl14,PlautMcClellandSeidenbergEtAl96}.  Likewise, empirical cognitive neuroscience is only beginning to uncover the relevant brain areas involved in these abilities \citep{ConstantinescuOReillyBehrens16,ParkMillerNiliEtAl20,SummerfieldLuyckxSheahan20,FranklandGreene20}.

The essential trick employed by the simple concatenation function is that arbitrary content can be routed into it and operated upon generically, independent of any details of the content.  The cognitive equivalent of this, which is widely recognized as critical for human systematic behavior \citep{FodorPylyshyn88} is a separation of \emph{structure} (i.e., the rule-processing system, akin to the function) from \emph{content}.  In language, syntax has traditionally been thought of as an example of content-independent structure, and Chomsky's famous example: ``colorless green ideas sleep furiously'' demonstrates that people can accurately judge grammatical correctness with arbitrary content.  Without adopting traditional abstract symbolic syntactic frameworks, understanding how something like content-independent syntactic structure can be learned and represented via known biological mechanisms is thus a critical step toward advancing our understanding of the neural basis of our signature human abilities.

In this paper, we review cognitive and neural evidence consistent with the separation between structure and content across various cognitive domains, and recent neural network models that demonstrate how such a separation might emerge through a combination of architectural biases and learning, producing significantly greater systematicity.  Critically, structure and content also need to interact and be integrated in various ways, and certain brain areas appear to be specialized for this integration, so as is often the case, the brain is likely to apply multiple representational strategies in parallel.

\section{Structure and Content in Cognitive Models}

The FINST (fingers of instantiation) framework of \citet{Pylyshyn89} provides an early, simple model for how structure can be represented independent of specific content and also maps well onto cognitive neuroscience data reviewed in the next section. The key idea is that abstract, content-independent, pointer-like ``fingers'' can index a small set (up to about 4) of different visual locations at a time.  These content-independent pointers were originally proposed to serve as place-holders for encoding relationships among items in a scene (e.g., \texttt{INSIDE(a,b)}).

The core ideas in the FINST framework can be extended to represent any kind of abstract structure that specifies the relationships between different indexed elements. For example, to describe the action of \emph{giving}, there are distinct functional roles (giver, recipient, item), and each of these could be represented using a separate FINST-like thematic role pointer, instead of using the semantic content specific to a particular situation (e.g., ``Radhika gave Charan a gift'').  Because \emph{giving} is encoded in terms of content-independent roles indexed by FINST-like pointers, these representations can automatically generalize to new contents, and additional inferences could be made about the properties and implications of the structural relationship (e.g., the giver no longer has the item, and may have some expectation of reciprocity depending on the nature of the transaction and relationship).  Encoding sufficiently rich structural knowledge to do so requires much more than FINST's, but encoding this knowledge in a relatively content-independent manner would clearly support systematic reasoning.

In short, FINST-like indexes provide a plausible attention-based neural mechanism for role-filler variable slots in the context of classical symbolic representational frameworks.  At a cognitive level, the kinds of elaborated structural representations and processes that have been studied include a range of different levels of complexity, from verb / action based elements as in the \emph{giving} example \citep{BoylanTrueswellThompson-Schill15} to more elaborated \emph{schemas} or \emph{scripts} describing longer sequences of events (e.g., the schema associated with a kid's birthday party), as we discuss below.  The ability to transfer structural knowledge across content domains is also central to established models of analogical reasoning \citep{GentnerHolyoak97} (e.g., the relational structure of \emph{orbiting} can be transferred from the planetary to atomic scale).

\section{Structure Representation in the Human Brain}

There is a well-established distinction between spatial and object processing in the human brain, which can be reframed as one example of how the brain separates structure and content via distinct, but interacting pathways.  Visual (and auditory) networks in the brain route sensory input into distinct dorsal and ventral stream pathways \citep{UngerleiderMishkin82}, with the ventral visual pathway extending from early visual cortex to inferotemporal (IT) cortex characterized as the \emph{What} pathway, specialized for object or scene recognition (i.e., visual \emph{content}).  The dorsal pathway through the parietal lobe is specialized for spatial \emph{Where} processing, based on extensive evidence that this pathway represents spatial and relational information in a relatively content-independent manner. 

The potential for the dorsal stream pathway to support systematic structure-sensitive processing was already well developed by \citet{Pylyshyn89}, and has been incorporated into psycholinguistic theories \citep{LandauJackendoff93,FranklandGreene20}.  Furthermore, there is substantial evidence that posterior parietal cortex encodes nonspatial structural information as well.  First, it is clear that the parietal lobe plays a critical role in the sensory guidance of action performance \citep{OrbanVanEssenVanduffel04}, as captured in the \emph{What} vs. \emph{How} ventral / dorsal framework  \citep{GoodaleMilner92}.  In humans, this action coding extends to the representation of verb-based argument structure (e.g., as in the \emph{giving} example above) in the inferior parietal area (angular gyrus), supporting the core of structure at the sentence level \citep{BoylanTrueswellThompson-Schill15,BinderDesai11,PalmerBonialHwang16}.  More abstract structural and linguistic concepts are thought to build directly upon these parietal action, space and time foundations \citep{LandauJackendoff93,FranklandGreene20}, including event representations in higher-order parietal areas as discussed below.

It is well established that parietal spatial and action representations anticipate the effects of eye, head, or body movements \citep{CavanaghHuntAfrazEtAl10} --- this suggests that acquisition of structure in the parietal lobe could be based on \emph{predictive learning} \citep{OReillyRussinZolfagharEtAl21}.  Specifically, learning driven by the difference between a predicted sensory outcome of an action and the subsequent sensory input can drive improved predictions and shape the formation of more abstract structural representations to more efficiently generate these predictions.  While it is fairly straightforward to learn to predict specific sensory outcomes from motor actions, an important area of current research is to determine the extent to which more abstract, structural representations can emerge, capturing the consistent, generalizable relationships that hold across a large number of such actions \citep{SummerfieldLuyckxSheahan20}.

Even if the dorsal-ventral streams separately encode structure and content in the brain, fundamental questions  remain about how these streams interact at the level of detailed neural mechanisms, to support systematic cognitive function.  A recent neural network model provides a useful example for how FINST-like attentional pointers can operate on newly-learned content information, in the context of separate structure and content representations \citep{RussinJoOReillyEtAl20}.  Unlike many existing models that have relied upon various hand-coded mechanisms to directly emulate programming-language like variable-binding functionality in neural hardware, this model learned entirely via error-driven learning, with only very broad, biologically plausible architectural constraints between two processing pathways.

\begin{figure}
  \centering\includegraphics[width=4in]{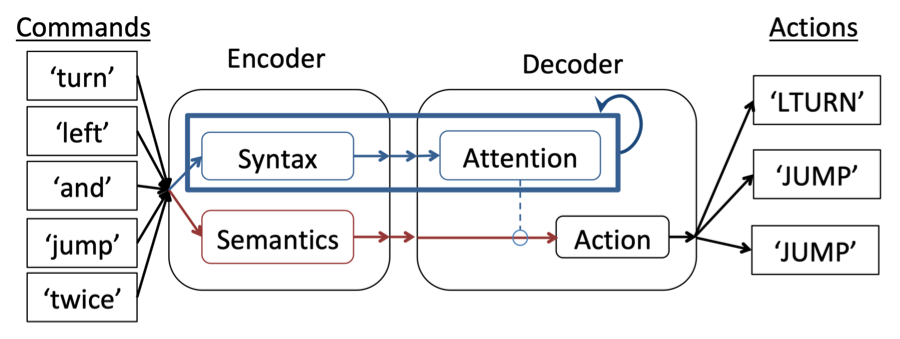}
  \caption{\footnotesize Diagram depicting a neural network model (Russin et al., 2020) with separate pathways for learning structure (syntax) and content (semantics). The structure pathway was recurrent (as shown by the blue self-connection) and had full access to the temporal ordering of the words in the instructions. However, it could only influence the output actions taken at each time step through an attention mechanism (indicated by the blue dotted line and open circle). The character of the representations in these different pathways facilitated systematic generalization, but was not built in, and emerged due to the architectural constraints imposed by the separation.}
  \label{fig.russin20}
\end{figure}

One such pathway had full access to the temporal ordering of words within a sentence, while the other was only able to process the single current word at any given time (Figure~\ref{fig.russin20}).  Furthermore, the temporal-order sensitive pathway could only influence the network output via attentional modulation of the other pathway (similar to the FINST-like attentional pointers).  Distinctions such as these could plausibly derive from evolved differences in the initial wiring of the neural architecture, providing affordances upon which subsequent learning operates.

With these constraints in place, the network learned syntax-like structural representations in the temporal-order pathway, and semantic-like content information in the other pathway, and in so doing, exhibited significantly more systematic behavior on challenging out-of-domain generalization tests than unstructured models \citep{LakeBaroni17}.  Specifically, when the model was trained on examples generated from a simple phrase structure grammar, the syntax-like pathway was able to direct attention to the proper sequence of content items based on commands like ``turn left twice,'' and generalize this to a novel verb, e.g., ``jump twice.''  The ability of attention-like mechanisms to drive this FINST-style indexing into novel content items is consistent with the central importance of such mechanisms in the neural basis of structure / content dynamics.  Furthermore, this model demonstrates how learned, distributed neural representations can take on a syntax-like role without traditional explicit symbolic grammar trees.  However,  more work is needed to understand the nature of structure representations in the brain supporting the more complex and sophisticated forms of human cognition, compared to these  simplified models.

Interestingly, a similar distinction between structure and content pathways was achieved in a fairly different way using a variable-binding system based on an explicit slot-filler (key-value) lookup table mechanism \citep{WebbSinhaCohen21}.  In this model, the structure processing pathway learned to control this lookup table independent of the specific content stored there, while the content pathway learned representations of individual items. This architectural structure / content separation enabled the model to exhibit much more systematic behavior than comparison models without such a separation.  Other related architectures are discussed in \citet{AkyurekAndreas21}.

While these models provide important initial demonstrations of how a separation between structure and content can support systematicity, it is critical to appreciate that many real-world situations also require strong interactions between the two. This has perhaps been most extensively studied in the case of pronouncing English words: as every English-as-a-second-language learner knows too well, the number of exceptions and sub-regularities are mind-spinning, and monolithic neural network pathways may provide the most appropriate mechanism for learning these mappings \citep{PlautMcClellandSeidenbergEtAl96}.  Likewise, the difference in parsing ``Fruit flies like a banana'' vs. ``Time flies like an arrow'' requires an interaction between syntax and semantics. In the visual domain, these interactions can be illustrated in the case of viewing a scene of a kitchen, where incoming content information may activate structural representations of typical kitchens, which then guide visual attention and semantic processing \citep{HayesHenderson19}.  In other words, human cognition likely represents a combination of separated and interacting streams \citep{OReillyPetrovCohenEtAl14,FranklinNormanRanganathEtAl20}.

\section{Binding of Structure and Content in the Medial Temporal Lobes}

\begin{figure}
  \centering\includegraphics[width=4in]{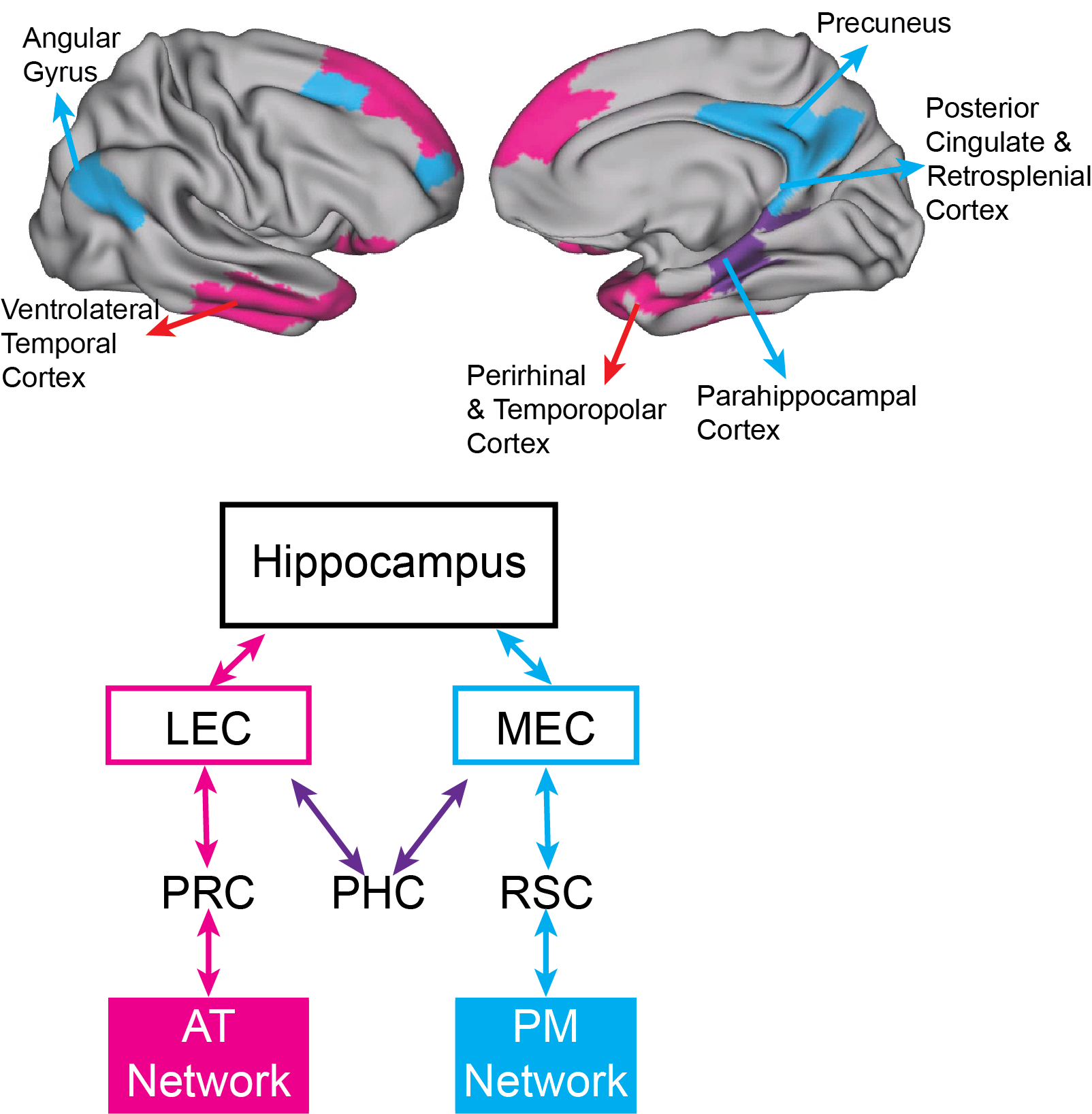}
  \caption{\footnotesize Two cortico-hippocampal networks. Top: Surface renderings depict the neocortical regions in the Posterior Medial (PM) Network (cyan), implicated in schema representation, and Anterior Temporal (AT) Network (magenta), implicated in representations of people and things. Bottom: A diagram schematically depicts the pathways by which information from the AT and PM Networks is directed to parallel pathways in the medial temporal lobes before fully converging in the hippocampus.  PRC = Perirhinal cortex; PHC = Parahippocampal cortex; RSC = Retrosplenial cortex; LEC = Lateral entorhinal cortex; MEC = Medial entorhinal cortex.}
  \label{fig.pmat}
\end{figure}

So far, we have used the classic model of the dorsal and ventral streams as one example of the separation of structure and content, but structure exists across multiple scales. For example, structural knowledge is required to parse and comprehend individual sentences, but comprehending a sentence within the context of a larger text passage often requires a mental representation of the structure of events depicted in that passage. Recent fMRI research has uncovered a candidate set of brain regions that may represent the abstract structure of events --- the \emph{Posterior Medial} (PM) network --- and regions that may represent the  characteristics of the specific people, places and things that serve as event content --- the \emph{Anterior Temporal} (AT) network (Figure~\ref{fig.pmat}). The PM and AT networks can be considered as anatomically higher up from the classic dorsal (PM) and ventral (AT) streams \citep{KravitzSaleemBakerEtAl11,RanganathRitchey12}. Although these networks are distinct, the hippocampus is in a key position to integrate information across the two networks.  AT network regions project predominantly to the lateral entorhinal cortex (LEC), and PM network regions mostly project to the medial entorhinal cortex (MEC).  The hippocampus proper then binds information from these streams together into unitary episodic memories \citep{RanganathRitchey12}.  Considerable evidence suggests that the PM and AT networks differentially represent information about structure and content, at higher, more complex levels and over longer spatiotemporal scales than the more posterior dorsal and ventral streams.  For example, the PM network encodes the structure of events and spatial contexts and participates in the reconstruction of past events and the simulation of future events \citep{RanganathRitchey12}, and is also central for discourse comprehension \citep{Martin-LoechesCasadoHernandez-TamamesEtAl08}.  

We can think of the hippocampus, PM, and AT networks as complementary learning systems \citep{RussinZolfagharParkEtAl21}, where the PM and AT networks are specialized for slow, integrative learning leading to highly generalizable representations, complementing the rapid episodic memory supported by the hippocampus.  By capturing states of PM and AT network regions at critical moments, the hippocampus is in a unique position to rapidly form memories that combine information about structure and content that is encountered during a specific period of time (i.e., an \emph{episode}).  When encountering a similar item or situation later on, hippocampal pattern completion can lead to reinstatement of previous states in the PM and AT networks, thereby facilitating the generation of a mental model of the current situation in a new or ambiguous context \citep{FranklinNormanRanganathEtAl20}.

Although they embody principles that are largely consistent with the ones proposed here, other computational models \citep{StachenfeldBotvinickGershman17,WhittingtonMullerMarkEtAl20} have placed more emphasis on the role of the hexagonally-tiled \emph{grid cells} in the MEC, which are thought to represent the topological structure of spatial environments, and hippocampal \emph{place cells}, which are thought to encode specific locations within an environment.  These models suggest that the MEC and hippocampus can represent any kind of task in a 2D state-space, consistent with fMRI studies showing MEC activity patterns characteristic of grid cells during the performance of novel tasks \citep{ConstantinescuOReillyBehrens16}. Notably, however, these and other studies of humans suggest that grid cells and grid cell-like fMRI activity patterns are evident throughout the PM network and in the medial prefrontal cortex \citep{JacobsWeidemannMillerEtAl13}.

Further work is needed to disentangle the unique contributions of the MEC relative to earlier parietal-lobe pathways in structure representation.  One possibility is that the use of structural knowledge to guide flexible behavior is driven by the PM network, and the MEC plays an important role in the use of episodic memory to support rapid structure learning by the PM network. For instance, it is possible that the MEC grid cell network might perform an initial step of \emph{pattern separation} on inputs from the PM network, moving them apart in representational space and thus making them easier to learn without interference (Frankland et al, in prep).  This is consistent with results from lesion studies suggesting that hippocampal or MEC lesions can significantly impair spatial learning, while generally sparing navigation in familiar environments  \citep{KolarikShahlaieHassanEtAl16,HalesSchlesigerLeutgebEtAl14}.  

\section{Conclusions}

There is good reason to believe that the architecture of the neocortex is optimized to support systematic and generative behavior by bifurcating sensory information into separate structure and content pathways along the dorsal and ventral streams. The hierarchically organized pathways through the parietal lobe generate different levels of representations that encode structural relationships reliably associated with different actions and events.  Such representations capture the general ``logic'' of actions and events: how things move, transform, and interact over space and time, in a way that can be readily applied to novel content. Significant work remains to be done in order to establish the nature of structure representations. By understanding the evolution of learning and dynamics across interacting brain networks, we can make significant progress toward understanding the nature of the representations that give rise to our uniquely human intelligence.

\section{Recommended Readings}

Pylyshyn, Z. (1989) (See References).  An original proposal for a structure / content separation mechanism based on attention, that we think may still have currency in neural network models of the brain.

Kravitz, D. J., Saleem, K. S., Baker, C. I., \& Mishkin, M. (2011) (See References).  Integrated review of three parietal lobe pathways supporting looking, reaching, and navigating.

Summerfield, C., Luyckx, F., \& Sheahan, H. (2020) (See References).  Recent synthetic review of a wide range of data and theory supporting the idea that the parietal lobe represents content-independent structure.

Whittington, J. C. R., et al., (2020) (See References). Computational model implementing structure / content separation in the hippocampal system, with nice illustrations of different forms of structure and how they can support systematic behavior in different domains.

% bibexport -o struct_curdir21.bib struct_curdir21_sub2.aux
%\bibliography{ccnlab}
\bibliography{struct_curdir21}

\begin{thebibliography}{}

\bibitem [\protect \citeauthoryear {%
Aky{\"u}rek%
\ \BBA {} Andreas%
}{%
Aky{\"u}rek%
\ \BBA {} Andreas%
}{%
{\protect \APACyear {2021}}%
}]{%
AkyurekAndreas21}
\APACinsertmetastar {%
AkyurekAndreas21}%
\begin{APACrefauthors}%
Aky{\"u}rek, E.%
\BCBT {}\ \BBA {} Andreas, J.%
\end{APACrefauthors}%
\unskip\
\newblock
\APACrefYearMonthDay{2021}{{\APACmonth{06}}}{}.
\newblock
{\BBOQ}\APACrefatitle {Lexicon {{Learning}} for {{Few}}-{{Shot Neural Sequence
  Modeling}}} {Lexicon {{Learning}} for {{Few}}-{{Shot Neural Sequence
  Modeling}}}.{\BBCQ}
\newblock
\APACjournalVolNumPages{arXiv:2106.03993 [cs]}{}{}{}.
\newblock
\begin{APACrefURL} \url{http://arxiv.org/abs/2106.03993} \end{APACrefURL}
\PrintBackRefs{\CurrentBib}

\bibitem [\protect \citeauthoryear {%
Binder%
\ \BBA {} Desai%
}{%
Binder%
\ \BBA {} Desai%
}{%
{\protect \APACyear {2011}}%
}]{%
BinderDesai11}
\APACinsertmetastar {%
BinderDesai11}%
\begin{APACrefauthors}%
Binder, J\BPBI R.%
\BCBT {}\ \BBA {} Desai, R\BPBI H.%
\end{APACrefauthors}%
\unskip\
\newblock
\APACrefYearMonthDay{2011}{{\APACmonth{11}}}{}.
\newblock
{\BBOQ}\APACrefatitle {The Neurobiology of Semantic Memory} {The neurobiology
  of semantic memory}.{\BBCQ}
\newblock
\APACjournalVolNumPages{Trends in Cognitive Sciences}{15}{11}{527--536}.
\newblock
\begin{APACrefDOI} \doi{10.1016/j.tics.2011.10.001} \end{APACrefDOI}
\PrintBackRefs{\CurrentBib}

\bibitem [\protect \citeauthoryear {%
Boylan%
, Trueswell%
\BCBL {}\ \BBA {} {Thompson-Schill}%
}{%
Boylan%
\ \protect \BOthers {.}}{%
{\protect \APACyear {2015}}%
}]{%
BoylanTrueswellThompson-Schill15}
\APACinsertmetastar {%
BoylanTrueswellThompson-Schill15}%
\begin{APACrefauthors}%
Boylan, C.%
, Trueswell, J\BPBI C.%
\BCBL {}\ \BBA {} {Thompson-Schill}, S\BPBI L.%
\end{APACrefauthors}%
\unskip\
\newblock
\APACrefYearMonthDay{2015}{{\APACmonth{11}}}{}.
\newblock
{\BBOQ}\APACrefatitle {Compositionality and the Angular Gyrus: {{A}}
  Multi-Voxel Similarity Analysis of the Semantic Composition of Nouns and
  Verbs} {Compositionality and the angular gyrus: {{A}} multi-voxel similarity
  analysis of the semantic composition of nouns and verbs}.{\BBCQ}
\newblock
\APACjournalVolNumPages{Neuropsychologia}{78}{}{130--141}.
\newblock
\begin{APACrefDOI} \doi{10.1016/j.neuropsychologia.2015.10.007}
  \end{APACrefDOI}
\PrintBackRefs{\CurrentBib}

\bibitem [\protect \citeauthoryear {%
Cavanagh%
, Hunt%
, Afraz%
\BCBL {}\ \BBA {} Rolfs%
}{%
Cavanagh%
\ \protect \BOthers {.}}{%
{\protect \APACyear {2010}}%
}]{%
CavanaghHuntAfrazEtAl10}
\APACinsertmetastar {%
CavanaghHuntAfrazEtAl10}%
\begin{APACrefauthors}%
Cavanagh, P.%
, Hunt, A\BPBI R.%
, Afraz, A.%
\BCBL {}\ \BBA {} Rolfs, M.%
\end{APACrefauthors}%
\unskip\
\newblock
\APACrefYearMonthDay{2010}{{\APACmonth{04}}}{}.
\newblock
{\BBOQ}\APACrefatitle {Visual Stability Based on Remapping of Attention
  Pointers} {Visual stability based on remapping of attention pointers}.{\BBCQ}
\newblock
\APACjournalVolNumPages{Trends in Cognitive Sciences}{14}{4}{147--153}.
\newblock
\begin{APACrefDOI} \doi{10.1016/j.tics.2010.01.007} \end{APACrefDOI}
\PrintBackRefs{\CurrentBib}

\bibitem [\protect \citeauthoryear {%
Constantinescu%
, O'Reilly%
\BCBL {}\ \BBA {} Behrens%
}{%
Constantinescu%
\ \protect \BOthers {.}}{%
{\protect \APACyear {2016}}%
}]{%
ConstantinescuOReillyBehrens16}
\APACinsertmetastar {%
ConstantinescuOReillyBehrens16}%
\begin{APACrefauthors}%
Constantinescu, A\BPBI O.%
, O'Reilly, J\BPBI X.%
\BCBL {}\ \BBA {} Behrens, T\BPBI E\BPBI J.%
\end{APACrefauthors}%
\unskip\
\newblock
\APACrefYearMonthDay{2016}{{\APACmonth{06}}}{}.
\newblock
{\BBOQ}\APACrefatitle {Organizing Conceptual Knowledge in Humans with a
  Gridlike Code} {Organizing conceptual knowledge in humans with a gridlike
  code}.{\BBCQ}
\newblock
\APACjournalVolNumPages{Science}{352}{6292}{1464--1468}.
\newblock
\begin{APACrefDOI} \doi{10.1126/science.aaf0941} \end{APACrefDOI}
\PrintBackRefs{\CurrentBib}

\bibitem [\protect \citeauthoryear {%
Fodor%
\ \BBA {} Pylyshyn%
}{%
Fodor%
\ \BBA {} Pylyshyn%
}{%
{\protect \APACyear {1988}}%
}]{%
FodorPylyshyn88}
\APACinsertmetastar {%
FodorPylyshyn88}%
\begin{APACrefauthors}%
Fodor, J\BPBI A.%
\BCBT {}\ \BBA {} Pylyshyn, Z\BPBI W.%
\end{APACrefauthors}%
\unskip\
\newblock
\APACrefYearMonthDay{1988}{{\APACmonth{04}}}{}.
\newblock
{\BBOQ}\APACrefatitle {Connectionism and Cognitive Architecture: A Critical
  Analysis.} {Connectionism and cognitive architecture: A critical
  analysis.}{\BBCQ}
\newblock
\APACjournalVolNumPages{Cognition}{28}{1-2}{3--71}.
\newblock
\begin{APACrefURL} \url{http://www.ncbi.nlm.nih.gov/pubmed/2450716}
  \end{APACrefURL}
\PrintBackRefs{\CurrentBib}

\bibitem [\protect \citeauthoryear {%
Frankland%
\ \BBA {} Greene%
}{%
Frankland%
\ \BBA {} Greene%
}{%
{\protect \APACyear {2020}}%
}]{%
FranklandGreene20}
\APACinsertmetastar {%
FranklandGreene20}%
\begin{APACrefauthors}%
Frankland, S\BPBI M.%
\BCBT {}\ \BBA {} Greene, J\BPBI D.%
\end{APACrefauthors}%
\unskip\
\newblock
\APACrefYearMonthDay{2020}{{\APACmonth{01}}}{}.
\newblock
{\BBOQ}\APACrefatitle {Concepts and {{Compositionality}}: {{In Search}} of the
  {{Brain}}'s {{Language}} of {{Thought}}} {Concepts and {{Compositionality}}:
  {{In Search}} of the {{Brain}}'s {{Language}} of {{Thought}}}.{\BBCQ}
\newblock
\APACjournalVolNumPages{Annual Review of Psychology}{71}{1}{273--303}.
\newblock
\begin{APACrefDOI} \doi{10.1146/annurev-psych-122216-011829} \end{APACrefDOI}
\PrintBackRefs{\CurrentBib}

\bibitem [\protect \citeauthoryear {%
Franklin%
, Norman%
, Ranganath%
, Zacks%
\BCBL {}\ \BBA {} Gershman%
}{%
Franklin%
\ \protect \BOthers {.}}{%
{\protect \APACyear {2020}}%
}]{%
FranklinNormanRanganathEtAl20}
\APACinsertmetastar {%
FranklinNormanRanganathEtAl20}%
\begin{APACrefauthors}%
Franklin, N\BPBI T.%
, Norman, K\BPBI A.%
, Ranganath, C.%
, Zacks, J\BPBI M.%
\BCBL {}\ \BBA {} Gershman, S\BPBI J.%
\end{APACrefauthors}%
\unskip\
\newblock
\APACrefYearMonthDay{2020}{{\APACmonth{04}}}{}.
\newblock
{\BBOQ}\APACrefatitle {Structured {{Event Memory}}: {{A}} Neuro-Symbolic Model
  of Event Cognition} {Structured {{Event Memory}}: {{A}} neuro-symbolic model
  of event cognition}.{\BBCQ}
\newblock
\APACjournalVolNumPages{Psychological Review}{127}{3}{327--361}.
\newblock
\begin{APACrefDOI} \doi{10.1037/rev0000177} \end{APACrefDOI}
\PrintBackRefs{\CurrentBib}

\bibitem [\protect \citeauthoryear {%
Gentner%
\ \BBA {} Holyoak%
}{%
Gentner%
\ \BBA {} Holyoak%
}{%
{\protect \APACyear {1997}}%
}]{%
GentnerHolyoak97}
\APACinsertmetastar {%
GentnerHolyoak97}%
\begin{APACrefauthors}%
Gentner, D.%
\BCBT {}\ \BBA {} Holyoak, K\BPBI J.%
\end{APACrefauthors}%
\unskip\
\newblock
\APACrefYearMonthDay{1997}{{\APACmonth{01}}}{}.
\newblock
{\BBOQ}\APACrefatitle {Reasoning and Learning by Analogy: {{Introduction}}}
  {Reasoning and learning by analogy: {{Introduction}}}.{\BBCQ}
\newblock
\APACjournalVolNumPages{American Psychologist}{52}{}{32--34}.
\PrintBackRefs{\CurrentBib}

\bibitem [\protect \citeauthoryear {%
Goodale%
\ \BBA {} Milner%
}{%
Goodale%
\ \BBA {} Milner%
}{%
{\protect \APACyear {1992}}%
}]{%
GoodaleMilner92}
\APACinsertmetastar {%
GoodaleMilner92}%
\begin{APACrefauthors}%
Goodale, M\BPBI A.%
\BCBT {}\ \BBA {} Milner, A\BPBI D.%
\end{APACrefauthors}%
\unskip\
\newblock
\APACrefYearMonthDay{1992}{{\APACmonth{01}}}{}.
\newblock
{\BBOQ}\APACrefatitle {Separate Visual Pathways for Perception and Action}
  {Separate visual pathways for perception and action}.{\BBCQ}
\newblock
\APACjournalVolNumPages{Trends in Neurosciences}{15}{1}{20--25}.
\PrintBackRefs{\CurrentBib}

\bibitem [\protect \citeauthoryear {%
Hales%
\ \protect \BOthers {.}}{%
Hales%
\ \protect \BOthers {.}}{%
{\protect \APACyear {2014}}%
}]{%
HalesSchlesigerLeutgebEtAl14}
\APACinsertmetastar {%
HalesSchlesigerLeutgebEtAl14}%
\begin{APACrefauthors}%
Hales, J\BPBI B.%
, Schlesiger, M\BPBI I.%
, Leutgeb, J\BPBI K.%
, Squire, L\BPBI R.%
, Leutgeb, S.%
\BCBL {}\ \BBA {} Clark, R\BPBI E.%
\end{APACrefauthors}%
\unskip\
\newblock
\APACrefYearMonthDay{2014}{{\APACmonth{11}}}{}.
\newblock
{\BBOQ}\APACrefatitle {Medial {{Entorhinal Cortex Lesions Only Partially
  Disrupt Hippocampal Place Cells}} and {{Hippocampus}}-{{Dependent Place
  Memory}}} {Medial {{Entorhinal Cortex Lesions Only Partially Disrupt
  Hippocampal Place Cells}} and {{Hippocampus}}-{{Dependent Place
  Memory}}}.{\BBCQ}
\newblock
\APACjournalVolNumPages{Cell Reports}{9}{3}{893--901}.
\newblock
\begin{APACrefDOI} \doi{10.1016/j.celrep.2014.10.009} \end{APACrefDOI}
\PrintBackRefs{\CurrentBib}

\bibitem [\protect \citeauthoryear {%
Hayes%
\ \BBA {} Henderson%
}{%
Hayes%
\ \BBA {} Henderson%
}{%
{\protect \APACyear {2019}}%
}]{%
HayesHenderson19}
\APACinsertmetastar {%
HayesHenderson19}%
\begin{APACrefauthors}%
Hayes, T\BPBI R.%
\BCBT {}\ \BBA {} Henderson, J\BPBI M.%
\end{APACrefauthors}%
\unskip\
\newblock
\APACrefYearMonthDay{2019}{{\APACmonth{10}}}{}.
\newblock
{\BBOQ}\APACrefatitle {Scene Semantics Involuntarily Guide Attention during
  Visual Search} {Scene semantics involuntarily guide attention during visual
  search}.{\BBCQ}
\newblock
\APACjournalVolNumPages{Psychonomic Bulletin \& Review}{26}{5}{1683--1689}.
\newblock
\begin{APACrefDOI} \doi{10.3758/s13423-019-01642-5} \end{APACrefDOI}
\PrintBackRefs{\CurrentBib}

\bibitem [\protect \citeauthoryear {%
Jacobs%
\ \protect \BOthers {.}}{%
Jacobs%
\ \protect \BOthers {.}}{%
{\protect \APACyear {2013}}%
}]{%
JacobsWeidemannMillerEtAl13}
\APACinsertmetastar {%
JacobsWeidemannMillerEtAl13}%
\begin{APACrefauthors}%
Jacobs, J.%
, Weidemann, C\BPBI T.%
, Miller, J\BPBI F.%
, Solway, A.%
, Burke, J\BPBI F.%
, Wei, X\BHBI X.%
\BDBL {}Kahana, M\BPBI J.%
\end{APACrefauthors}%
\unskip\
\newblock
\APACrefYearMonthDay{2013}{{\APACmonth{09}}}{}.
\newblock
{\BBOQ}\APACrefatitle {Direct Recordings of Grid-like Neuronal Activity in
  Human Spatial Navigation} {Direct recordings of grid-like neuronal activity
  in human spatial navigation}.{\BBCQ}
\newblock
\APACjournalVolNumPages{Nature Neuroscience}{16}{9}{1188--1190}.
\newblock
\begin{APACrefDOI} \doi{10.1038/nn.3466} \end{APACrefDOI}
\PrintBackRefs{\CurrentBib}

\bibitem [\protect \citeauthoryear {%
Kolarik%
\ \protect \BOthers {.}}{%
Kolarik%
\ \protect \BOthers {.}}{%
{\protect \APACyear {2016}}%
}]{%
KolarikShahlaieHassanEtAl16}
\APACinsertmetastar {%
KolarikShahlaieHassanEtAl16}%
\begin{APACrefauthors}%
Kolarik, B\BPBI S.%
, Shahlaie, K.%
, Hassan, A.%
, Borders, A\BPBI A.%
, Kaufman, K\BPBI C.%
, Gurkoff, G.%
\BDBL {}Ekstrom, A\BPBI D.%
\end{APACrefauthors}%
\unskip\
\newblock
\APACrefYearMonthDay{2016}{{\APACmonth{01}}}{}.
\newblock
{\BBOQ}\APACrefatitle {Impairments in Precision, Rather than Spatial Strategy,
  Characterize Performance on the Virtual {{Morris Water Maze}}: {{A}} Case
  Study} {Impairments in precision, rather than spatial strategy, characterize
  performance on the virtual {{Morris Water Maze}}: {{A}} case study}.{\BBCQ}
\newblock
\APACjournalVolNumPages{Neuropsychologia}{80}{}{90--101}.
\newblock
\begin{APACrefDOI} \doi{10.1016/j.neuropsychologia.2015.11.013}
  \end{APACrefDOI}
\PrintBackRefs{\CurrentBib}

\bibitem [\protect \citeauthoryear {%
Kravitz%
, Saleem%
, Baker%
\BCBL {}\ \BBA {} Mishkin%
}{%
Kravitz%
\ \protect \BOthers {.}}{%
{\protect \APACyear {2011}}%
}]{%
KravitzSaleemBakerEtAl11}
\APACinsertmetastar {%
KravitzSaleemBakerEtAl11}%
\begin{APACrefauthors}%
Kravitz, D\BPBI J.%
, Saleem, K\BPBI S.%
, Baker, C\BPBI I.%
\BCBL {}\ \BBA {} Mishkin, M.%
\end{APACrefauthors}%
\unskip\
\newblock
\APACrefYearMonthDay{2011}{{\APACmonth{04}}}{}.
\newblock
{\BBOQ}\APACrefatitle {A New Neural Framework for Visuospatial Processing} {A
  new neural framework for visuospatial processing}.{\BBCQ}
\newblock
\APACjournalVolNumPages{Nature Reviews Neuroscience}{12}{4}{217--230}.
\newblock
\begin{APACrefDOI} \doi{10.1038/nrn3008} \end{APACrefDOI}
\PrintBackRefs{\CurrentBib}

\bibitem [\protect \citeauthoryear {%
Lake%
\ \BBA {} Baroni%
}{%
Lake%
\ \BBA {} Baroni%
}{%
{\protect \APACyear {2017}}%
}]{%
LakeBaroni17}
\APACinsertmetastar {%
LakeBaroni17}%
\begin{APACrefauthors}%
Lake, B\BPBI M.%
\BCBT {}\ \BBA {} Baroni, M.%
\end{APACrefauthors}%
\unskip\
\newblock
\APACrefYearMonthDay{2017}{{\APACmonth{10}}}{}.
\newblock
{\BBOQ}\APACrefatitle {Generalization without Systematicity: {{On}} the
  Compositional Skills of Sequence-to-Sequence Recurrent Networks}
  {Generalization without systematicity: {{On}} the compositional skills of
  sequence-to-sequence recurrent networks}.{\BBCQ}
\newblock
\APACjournalVolNumPages{arXiv:1711.00350 [cs]}{}{}{}.
\newblock
\begin{APACrefURL} \url{http://arxiv.org/abs/1711.00350} \end{APACrefURL}
\PrintBackRefs{\CurrentBib}

\bibitem [\protect \citeauthoryear {%
Landau%
\ \BBA {} Jackendoff%
}{%
Landau%
\ \BBA {} Jackendoff%
}{%
{\protect \APACyear {1993}}%
}]{%
LandauJackendoff93}
\APACinsertmetastar {%
LandauJackendoff93}%
\begin{APACrefauthors}%
Landau, B.%
\BCBT {}\ \BBA {} Jackendoff, R.%
\end{APACrefauthors}%
\unskip\
\newblock
\APACrefYearMonthDay{1993}{}{}.
\newblock
{\BBOQ}\APACrefatitle {"{{What}}" and "Where" in Spatial Language and Spatial
  Cognition} {"{{What}}" and "where" in spatial language and spatial
  cognition}.{\BBCQ}
\newblock
\APACjournalVolNumPages{Behavioral and Brain Sciences}{16}{2}{217--265}.
\newblock
\begin{APACrefDOI} \doi{10.1017/S0140525X00029733} \end{APACrefDOI}
\PrintBackRefs{\CurrentBib}

\bibitem [\protect \citeauthoryear {%
Marcus%
}{%
Marcus%
}{%
{\protect \APACyear {2018}}%
}]{%
Marcus18}
\APACinsertmetastar {%
Marcus18}%
\begin{APACrefauthors}%
Marcus, G.%
\end{APACrefauthors}%
\unskip\
\newblock
\APACrefYearMonthDay{2018}{{\APACmonth{01}}}{}.
\newblock
{\BBOQ}\APACrefatitle {Deep {{Learning}}: {{A Critical Appraisal}}} {Deep
  {{Learning}}: {{A Critical Appraisal}}}.{\BBCQ}
\newblock
\APACjournalVolNumPages{arXiv:1801.00631 [cs, stat]}{}{}{}.
\newblock
\begin{APACrefURL} \url{http://arxiv.org/abs/1801.00631} \end{APACrefURL}
\PrintBackRefs{\CurrentBib}

\bibitem [\protect \citeauthoryear {%
{Mart{\'i}n-Loeches}%
, Casado%
, {Hern{\'a}ndez-Tamames}%
\BCBL {}\ \BBA {} {\'A}lvarez-Linera%
}{%
{Mart{\'i}n-Loeches}%
\ \protect \BOthers {.}}{%
{\protect \APACyear {2008}}%
}]{%
Martin-LoechesCasadoHernandez-TamamesEtAl08}
\APACinsertmetastar {%
Martin-LoechesCasadoHernandez-TamamesEtAl08}%
\begin{APACrefauthors}%
{Mart{\'i}n-Loeches}, M.%
, Casado, P.%
, {Hern{\'a}ndez-Tamames}, J\BPBI A.%
\BCBL {}\ \BBA {} {\'A}lvarez-Linera, J.%
\end{APACrefauthors}%
\unskip\
\newblock
\APACrefYearMonthDay{2008}{{\APACmonth{06}}}{}.
\newblock
{\BBOQ}\APACrefatitle {Brain Activation in Discourse Comprehension: {{A}} 3t
  {{fMRI}} Study} {Brain activation in discourse comprehension: {{A}} 3t
  {{fMRI}} study}.{\BBCQ}
\newblock
\APACjournalVolNumPages{NeuroImage}{41}{2}{614--622}.
\newblock
\begin{APACrefDOI} \doi{10.1016/j.neuroimage.2008.02.047} \end{APACrefDOI}
\PrintBackRefs{\CurrentBib}

\bibitem [\protect \citeauthoryear {%
Orban%
, Van~Essen%
\BCBL {}\ \BBA {} Vanduffel%
}{%
Orban%
\ \protect \BOthers {.}}{%
{\protect \APACyear {2004}}%
}]{%
OrbanVanEssenVanduffel04}
\APACinsertmetastar {%
OrbanVanEssenVanduffel04}%
\begin{APACrefauthors}%
Orban, G\BPBI A.%
, Van~Essen, D.%
\BCBL {}\ \BBA {} Vanduffel, W.%
\end{APACrefauthors}%
\unskip\
\newblock
\APACrefYearMonthDay{2004}{{\APACmonth{01}}}{}.
\newblock
{\BBOQ}\APACrefatitle {Comparative Mapping of Higher Visual Areas in Monkeys
  and Humans} {Comparative mapping of higher visual areas in monkeys and
  humans}.{\BBCQ}
\newblock
\APACjournalVolNumPages{Trends in Cognitive Sciences.}{8}{7}{315--324}.
\PrintBackRefs{\CurrentBib}

\bibitem [\protect \citeauthoryear {%
O'Reilly%
\ \protect \BOthers {.}}{%
O'Reilly%
\ \protect \BOthers {.}}{%
{\protect \APACyear {2014}}%
}]{%
OReillyPetrovCohenEtAl14}
\APACinsertmetastar {%
OReillyPetrovCohenEtAl14}%
\begin{APACrefauthors}%
O'Reilly, R\BPBI C.%
, Petrov, A\BPBI A.%
, Cohen, J\BPBI D.%
, Lebiere, C\BPBI J.%
, Herd, S\BPBI A.%
\BCBL {}\ \BBA {} Kriete, T.%
\end{APACrefauthors}%
\unskip\
\newblock
\APACrefYearMonthDay{2014}{}{}.
\newblock
{\BBOQ}\APACrefatitle {How {{Limited Systematicity Emerges}}: {{A Computational
  Cognitive Neuroscience Approach}}} {How {{Limited Systematicity Emerges}}:
  {{A Computational Cognitive Neuroscience Approach}}}.{\BBCQ}
\newblock
\BIn{} I\BPBI P.~Calvo\ \BBA {} J.~Symons\ (\BEDS), \APACrefbtitle {The
  Architecture of Cognition: {{Rethinking Fodor}} and {{Pylyshyn}}{$^1$}s
  {{Systematicity Challenge}}.} {The architecture of cognition: {{Rethinking
  Fodor}} and {{Pylyshyn}}{$^1$}s {{Systematicity Challenge}}.}
\newblock
\APACaddressPublisher{{Cambridge, MA}}{{MIT Press}}.
\PrintBackRefs{\CurrentBib}

\bibitem [\protect \citeauthoryear {%
O'Reilly%
, Russin%
, Zolfaghar%
\BCBL {}\ \BBA {} Rohrlich%
}{%
O'Reilly%
\ \protect \BOthers {.}}{%
{\protect \APACyear {2021}}%
}]{%
OReillyRussinZolfagharEtAl21}
\APACinsertmetastar {%
OReillyRussinZolfagharEtAl21}%
\begin{APACrefauthors}%
O'Reilly, R\BPBI C.%
, Russin, J\BPBI L.%
, Zolfaghar, M.%
\BCBL {}\ \BBA {} Rohrlich, J.%
\end{APACrefauthors}%
\unskip\
\newblock
\APACrefYearMonthDay{2021}{}{}.
\newblock
{\BBOQ}\APACrefatitle {Deep Predictive Learning in Neocortex and Pulvinar}
  {Deep predictive learning in neocortex and pulvinar}.{\BBCQ}
\newblock
\APACjournalVolNumPages{Journal of Cognitive Neuroscience}{33}{6}{1158--1196}.
\PrintBackRefs{\CurrentBib}

\bibitem [\protect \citeauthoryear {%
Palmer%
, Bonial%
\BCBL {}\ \BBA {} Hwang%
}{%
Palmer%
\ \protect \BOthers {.}}{%
{\protect \APACyear {2016}}%
}]{%
PalmerBonialHwang16}
\APACinsertmetastar {%
PalmerBonialHwang16}%
\begin{APACrefauthors}%
Palmer, M.%
, Bonial, C.%
\BCBL {}\ \BBA {} Hwang, J\BPBI D.%
\end{APACrefauthors}%
\unskip\
\newblock
\APACrefYearMonthDay{2016}{{\APACmonth{11}}}{}.
\newblock
{\BBOQ}\APACrefatitle {{{VerbNet}}: {{Capturing English}} Verb Behavior,
  Meaning and Usage} {{{VerbNet}}: {{Capturing English}} verb behavior, meaning
  and usage}.{\BBCQ}
\newblock
\BIn{} S\BPBI E\BPBI F.~Chipman\ (\BED), \APACrefbtitle {The {{Oxford
  Handbook}} of {{Cognitive Science}}} {The {{Oxford Handbook}} of {{Cognitive
  Science}}}\ (\BPGS\ 315--336).
\newblock
\APACaddressPublisher{}{{Oxford University Press}}.
\PrintBackRefs{\CurrentBib}

\bibitem [\protect \citeauthoryear {%
Park%
, Miller%
, Nili%
, Ranganath%
\BCBL {}\ \BBA {} Boorman%
}{%
Park%
\ \protect \BOthers {.}}{%
{\protect \APACyear {2020}}%
}]{%
ParkMillerNiliEtAl20}
\APACinsertmetastar {%
ParkMillerNiliEtAl20}%
\begin{APACrefauthors}%
Park, S\BPBI A.%
, Miller, D\BPBI S.%
, Nili, H.%
, Ranganath, C.%
\BCBL {}\ \BBA {} Boorman, E\BPBI D.%
\end{APACrefauthors}%
\unskip\
\newblock
\APACrefYearMonthDay{2020}{{\APACmonth{07}}}{}.
\newblock
{\BBOQ}\APACrefatitle {Map {{Making}}: {{Constructing}}, {{Combining}}, and
  {{Inferring}} on {{Abstract Cognitive Maps}}} {Map {{Making}}:
  {{Constructing}}, {{Combining}}, and {{Inferring}} on {{Abstract Cognitive
  Maps}}}.{\BBCQ}
\newblock
\APACjournalVolNumPages{Neuron}{}{}{}.
\newblock
\begin{APACrefDOI} \doi{10.1016/j.neuron.2020.06.030} \end{APACrefDOI}
\PrintBackRefs{\CurrentBib}

\bibitem [\protect \citeauthoryear {%
Plaut%
, McClelland%
, Seidenberg%
\BCBL {}\ \BBA {} Patterson%
}{%
Plaut%
\ \protect \BOthers {.}}{%
{\protect \APACyear {1996}}%
}]{%
PlautMcClellandSeidenbergEtAl96}
\APACinsertmetastar {%
PlautMcClellandSeidenbergEtAl96}%
\begin{APACrefauthors}%
Plaut, D\BPBI C.%
, McClelland, J\BPBI L.%
, Seidenberg, M\BPBI S.%
\BCBL {}\ \BBA {} Patterson, K.%
\end{APACrefauthors}%
\unskip\
\newblock
\APACrefYearMonthDay{1996}{{\APACmonth{07}}}{}.
\newblock
{\BBOQ}\APACrefatitle {Understanding Normal and Impaired Word Reading:
  Computational Principles in Quasi-Regular Domains.} {Understanding normal and
  impaired word reading: Computational principles in quasi-regular
  domains.}{\BBCQ}
\newblock
\APACjournalVolNumPages{Psychological review}{103}{}{56--115}.
\newblock
\begin{APACrefURL} \url{http://www.ncbi.nlm.nih.gov/pubmed/8650300}
  \end{APACrefURL}
\PrintBackRefs{\CurrentBib}

\bibitem [\protect \citeauthoryear {%
Pylyshyn%
}{%
Pylyshyn%
}{%
{\protect \APACyear {1989}}%
}]{%
Pylyshyn89}
\APACinsertmetastar {%
Pylyshyn89}%
\begin{APACrefauthors}%
Pylyshyn, Z.%
\end{APACrefauthors}%
\unskip\
\newblock
\APACrefYearMonthDay{1989}{{\APACmonth{06}}}{}.
\newblock
{\BBOQ}\APACrefatitle {The Role of Location Indexes in Spatial Perception:
  {{A}} Sketch of the {{FINST}} Spatial-Index Model} {The role of location
  indexes in spatial perception: {{A}} sketch of the {{FINST}} spatial-index
  model}.{\BBCQ}
\newblock
\APACjournalVolNumPages{Cognition}{32}{1}{65--97}.
\newblock
\begin{APACrefDOI} \doi{10.1016/0010-0277(89)90014-0} \end{APACrefDOI}
\PrintBackRefs{\CurrentBib}

\bibitem [\protect \citeauthoryear {%
Ranganath%
\ \BBA {} Ritchey%
}{%
Ranganath%
\ \BBA {} Ritchey%
}{%
{\protect \APACyear {2012}}%
}]{%
RanganathRitchey12}
\APACinsertmetastar {%
RanganathRitchey12}%
\begin{APACrefauthors}%
Ranganath, C.%
\BCBT {}\ \BBA {} Ritchey, M.%
\end{APACrefauthors}%
\unskip\
\newblock
\APACrefYearMonthDay{2012}{{\APACmonth{10}}}{}.
\newblock
{\BBOQ}\APACrefatitle {Two Cortical Systems for Memory-Guided Behaviour.} {Two
  cortical systems for memory-guided behaviour.}{\BBCQ}
\newblock
\APACjournalVolNumPages{Nature Reviews Neuroscience}{13}{10}{713--726}.
\newblock
\begin{APACrefURL} \url{http://www.ncbi.nlm.nih.gov/pubmed/22992647}
  \end{APACrefURL}
\PrintBackRefs{\CurrentBib}

\bibitem [\protect \citeauthoryear {%
Russin%
, Jo%
, O'Reilly%
\BCBL {}\ \BBA {} Bengio%
}{%
Russin%
\ \protect \BOthers {.}}{%
{\protect \APACyear {2020}}%
}]{%
RussinJoOReillyEtAl20}
\APACinsertmetastar {%
RussinJoOReillyEtAl20}%
\begin{APACrefauthors}%
Russin, J\BPBI L.%
, Jo, J.%
, O'Reilly, R\BPBI C.%
\BCBL {}\ \BBA {} Bengio, Y.%
\end{APACrefauthors}%
\unskip\
\newblock
\APACrefYearMonthDay{2020}{}{}.
\newblock
{\BBOQ}\APACrefatitle {Systematicity in a {{Recurrent Neural Network}} by
  {{Factorizing Syntax}} and {{Semantics}}} {Systematicity in a {{Recurrent
  Neural Network}} by {{Factorizing Syntax}} and {{Semantics}}}.{\BBCQ}
\newblock
\BIn{} \APACrefbtitle {Proceedings for the 42nd {{Annual Meeting}} of the
  {{Cognitive Science Society}}} {Proceedings for the 42nd {{Annual Meeting}}
  of the {{Cognitive Science Society}}}\ (\BPG~7).
\PrintBackRefs{\CurrentBib}

\bibitem [\protect \citeauthoryear {%
Russin%
, Zolfaghar%
, Park%
, Boorman%
\BCBL {}\ \BBA {} O'Reilly%
}{%
Russin%
\ \protect \BOthers {.}}{%
{\protect \APACyear {2021}}%
}]{%
RussinZolfagharParkEtAl21}
\APACinsertmetastar {%
RussinZolfagharParkEtAl21}%
\begin{APACrefauthors}%
Russin, J\BPBI L.%
, Zolfaghar, M.%
, Park, S\BPBI A.%
, Boorman, E.%
\BCBL {}\ \BBA {} O'Reilly, R\BPBI C.%
\end{APACrefauthors}%
\unskip\
\newblock
\APACrefYearMonthDay{2021}{}{}.
\newblock
{\BBOQ}\APACrefatitle {Complementary {{Structure}}-{{Learning Neural Networks}}
  for {{Relational Reasoning}}} {Complementary {{Structure}}-{{Learning Neural
  Networks}} for {{Relational Reasoning}}}.{\BBCQ}
\newblock
\BIn{} \APACrefbtitle {Proceedings for the 43nd {{Annual Meeting}} of the
  {{Cognitive Science Society}}} {Proceedings for the 43nd {{Annual Meeting}}
  of the {{Cognitive Science Society}}}\ (\BPG~7).
\PrintBackRefs{\CurrentBib}

\bibitem [\protect \citeauthoryear {%
Stachenfeld%
, Botvinick%
\BCBL {}\ \BBA {} Gershman%
}{%
Stachenfeld%
\ \protect \BOthers {.}}{%
{\protect \APACyear {2017}}%
}]{%
StachenfeldBotvinickGershman17}
\APACinsertmetastar {%
StachenfeldBotvinickGershman17}%
\begin{APACrefauthors}%
Stachenfeld, K\BPBI L.%
, Botvinick, M\BPBI M.%
\BCBL {}\ \BBA {} Gershman, S\BPBI J.%
\end{APACrefauthors}%
\unskip\
\newblock
\APACrefYearMonthDay{2017}{{\APACmonth{11}}}{}.
\newblock
{\BBOQ}\APACrefatitle {The Hippocampus as a Predictive Map} {The hippocampus as
  a predictive map}.{\BBCQ}
\newblock
\APACjournalVolNumPages{Nature Neuroscience}{20}{11}{1643--1653}.
\newblock
\begin{APACrefDOI} \doi{10.1038/nn.4650} \end{APACrefDOI}
\PrintBackRefs{\CurrentBib}

\bibitem [\protect \citeauthoryear {%
Summerfield%
, Luyckx%
\BCBL {}\ \BBA {} Sheahan%
}{%
Summerfield%
\ \protect \BOthers {.}}{%
{\protect \APACyear {2020}}%
}]{%
SummerfieldLuyckxSheahan20}
\APACinsertmetastar {%
SummerfieldLuyckxSheahan20}%
\begin{APACrefauthors}%
Summerfield, C.%
, Luyckx, F.%
\BCBL {}\ \BBA {} Sheahan, H.%
\end{APACrefauthors}%
\unskip\
\newblock
\APACrefYearMonthDay{2020}{{\APACmonth{01}}}{}.
\newblock
{\BBOQ}\APACrefatitle {Structure Learning and the Posterior Parietal Cortex}
  {Structure learning and the posterior parietal cortex}.{\BBCQ}
\newblock
\APACjournalVolNumPages{Progress in Neurobiology}{184}{}{101717}.
\newblock
\begin{APACrefDOI} \doi{10.1016/j.pneurobio.2019.101717} \end{APACrefDOI}
\PrintBackRefs{\CurrentBib}

\bibitem [\protect \citeauthoryear {%
Ungerleider%
\ \BBA {} Mishkin%
}{%
Ungerleider%
\ \BBA {} Mishkin%
}{%
{\protect \APACyear {1982}}%
}]{%
UngerleiderMishkin82}
\APACinsertmetastar {%
UngerleiderMishkin82}%
\begin{APACrefauthors}%
Ungerleider, L\BPBI G.%
\BCBT {}\ \BBA {} Mishkin, M.%
\end{APACrefauthors}%
\unskip\
\newblock
\APACrefYearMonthDay{1982}{{\APACmonth{01}}}{}.
\newblock
{\BBOQ}\APACrefatitle {Two {{Cortical Visual Systems}}} {Two {{Cortical Visual
  Systems}}}.{\BBCQ}
\newblock
\BIn{} D\BPBI J.~Ingle, M\BPBI A.~Goodale\BCBL {}\ \BBA {} R\BPBI J\BPBI
  W.~Mansfield\ (\BEDS), \APACrefbtitle {The {{Analysis}} of {{Visual
  Behavior}}} {The {{Analysis}} of {{Visual Behavior}}}\ (\BPGS\ 549--586).
\newblock
\APACaddressPublisher{{Cambridge, MA}}{{MIT Press}}.
\PrintBackRefs{\CurrentBib}

\bibitem [\protect \citeauthoryear {%
Webb%
, Sinha%
\BCBL {}\ \BBA {} Cohen%
}{%
Webb%
\ \protect \BOthers {.}}{%
{\protect \APACyear {2021}}%
}]{%
WebbSinhaCohen21}
\APACinsertmetastar {%
WebbSinhaCohen21}%
\begin{APACrefauthors}%
Webb, T\BPBI W.%
, Sinha, I.%
\BCBL {}\ \BBA {} Cohen, J\BPBI D.%
\end{APACrefauthors}%
\unskip\
\newblock
\APACrefYearMonthDay{2021}{{\APACmonth{03}}}{}.
\newblock
{\BBOQ}\APACrefatitle {Emergent {{Symbols}} through {{Binding}} in {{External
  Memory}}} {Emergent {{Symbols}} through {{Binding}} in {{External
  Memory}}}.{\BBCQ}
\newblock
\APACjournalVolNumPages{ICLR 2021: Proceedings of the International Conference
  on Learning Representations}{}{}{}.
\newblock
\begin{APACrefURL} \url{http://arxiv.org/abs/2012.14601} \end{APACrefURL}
\PrintBackRefs{\CurrentBib}

\bibitem [\protect \citeauthoryear {%
Whittington%
\ \protect \BOthers {.}}{%
Whittington%
\ \protect \BOthers {.}}{%
{\protect \APACyear {2020}}%
}]{%
WhittingtonMullerMarkEtAl20}
\APACinsertmetastar {%
WhittingtonMullerMarkEtAl20}%
\begin{APACrefauthors}%
Whittington, J\BPBI C\BPBI R.%
, Muller, T\BPBI H.%
, Mark, S.%
, Chen, G.%
, Barry, C.%
, Burgess, N.%
\BCBL {}\ \BBA {} Behrens, T\BPBI E\BPBI J.%
\end{APACrefauthors}%
\unskip\
\newblock
\APACrefYearMonthDay{2020}{{\APACmonth{11}}}{}.
\newblock
{\BBOQ}\APACrefatitle {The {{Tolman}}-{{Eichenbaum Machine}}: {{Unifying}}
  Space and Relational Memory through Generalization in the Hippocampal
  Formation} {The {{Tolman}}-{{Eichenbaum Machine}}: {{Unifying}} space and
  relational memory through generalization in the hippocampal
  formation}.{\BBCQ}
\newblock
\APACjournalVolNumPages{Cell}{0}{0}{}.
\newblock
\begin{APACrefDOI} \doi{10.1016/j.cell.2020.10.024} \end{APACrefDOI}
\PrintBackRefs{\CurrentBib}

\end{thebibliography}

\end{document}